\documentclass[12pt]{article}
\pdfoutput =1
\usepackage{graphicx}
\DeclareGraphicsExtensions{.pdf}
\usepackage{float} 
\usepackage{amsmath}
\usepackage{amsfonts}   
\usepackage{amssymb}

\begin{document}
\date{}
\title{{\bf{\Large AdS/CFT superconductors with Power Maxwell electrodynamics: reminiscent of the Meissner effect}}}
\author{
 {\bf {\normalsize Dibakar Roychowdhury}$
$\thanks{E-mail: dibakar@bose.res.in, dibakarphys@gmail.com}}\\
 {\normalsize S.~N.~Bose National Centre for Basic Sciences,}
\\{\normalsize JD Block, Sector III, Salt Lake, Kolkata-700098, India}
\\[0.3cm]
}

\maketitle
\begin{abstract}
 Based on an analytic scheme and neglecting the back reaction effect several crucial properties of holographic $ s $- wave superconductors have been investigated in the presence of an external magnetic field in the background of a $ D $ dimensional Schwarzschild AdS space time. Inspired by low energy limit of heterotic string theory, in the present paper we replace the conventional Maxwell action by a Power Maxwell action.  Immersing the holographic superconductors in an external static magnetic field the spatially dependent condensate solutions have been obtained analytically. Interestingly enough it is observed that condensation can form only below a certain critical field strength ($ B_c $). Finally, and most importantly it is observed that the value of this critical field strength increases as the mass of the scalar particles gets higher, which indicates the onset of a harder condensation.
\end{abstract}

\section{Introduction}
 
 The AdS/CFT duality \cite{ref1}-\cite{ref2}, which provides an exact connection between the classical gravitational theory in $ (D+1) $ dimensions to that with the large $ N $ limit of certain classes of gauge theories in $ D $ dimensions is considered to be the most significant achievement of string theory so far. Due to its several remarkable features the theory has attained renewed attention for the past couple of years. One of the most significant achievement of this gauge/gravity correspondence \cite{ref3} is that it provides a surprising connection between the classical general relativity and other non gravitational areas of physics e,g; various condensed matter phenomena \cite{ref4}-\cite{ref5}. More specifically, exploiting the gauge/gravity duality one can use general relativity as a tool in order to describe various strongly correlated systems of condensed matter physics, including even the phenomena of superconductivity\cite{ref6}-\cite{ref7}.  
 
 It was Gubser who first argued that there could be a spontaneous $ U(1) $ symmetry breaking near the event horizon of a charged black hole if the black hole is sufficiently charged and minimally coupled to a complex scalar field \cite{ref8}-\cite{ref10}. In the language of gauge/gravity duality such a local symmetry breaking in the bulk theory corresponds to a global $ U(1) $ symmetry breaking in the dual field theory residing at the boundary. Spontaneous breaking of such a global $ U(1) $ symmetry basically triggers a superconductivity in the boundary field theory popularly known as holographic superconductivity. Since then this correspondence has been widely explored in order to understand several crucial properties of these holographic superconductors \cite{ref11}-\cite{Banerjee:2012vk}.
 
 One of the major characteristic properties of ordinary superconductors is that they expel magnetic fields as the temperature is lowered through the critical temperature ($ T_c $) \cite{ref33}-\cite{ref34}.  In the presence of an external magnetic field, ordinary superconductors may be classified into two categories, namely type I and type II. 
 On the other hand, using the AdS/CFT dictionary and mostly based on \textit{numerical techniques }it has been found that at low temperatures ($ T<T_c $) magnetic field expels the $ s $- wave condensate in ($ 2+1 $)  holographic superconductors along with the formation of vortices \cite{ref35}-\cite{ref39}. This behavior is similar to that is observed in ordinary type II superconductors.  In spite of all these attempts, there are indeed few significant issues which remain highly debatable and worthy of further investigations. These may be put as follows:
 (1) Until now all the attempts in order to investigate the behavior of condensate solutions in the presence of an external magnetic field has been carried out in the framework of usual Maxwell electrodynamics and particularly for ($ 2+1 $) dimensional superconductors. Therefore, it is an important question how the condensation behaves when higher curvature corrections are incorporated in the usual Maxwell action in general in any dimensions\footnote{Recently theories with non linear electrodynamics has attained renewed attention due to its natural emergence in the low energy limit of heterotic string theory. Moreover, non linear electrodynamics models provide interesting black hole solutions which besides satisfying the zeroth and first law of black hole mechanics also possesses various other appealing features, like regular black hole solutions \cite{ref40}-\cite{ref46}.}'\footnote{Although, for the past couple of years, studying holographic superconductors in the framework of non linear electrodynamics has been a popular topic of research \cite{ref27}-\cite{nref40}, still till date no analysis has been performed incorporating the effects of these non linear corrections to the $ s $- wave condensate in presence of external magnetic field. }.
 (2) Secondly, it will also be interesting to explore how the $ s $- wave condensate depends on the scalar mass ($ m $) when the magnetic field is turned on. This is indeed a crucial issue which has never been explored so far\footnote{In \cite{ref12} based on numerical techniques, the authors explored the effect of scalar mass on $ s $- wave condensate in the absence of any external magnetic field and within the framework of Maxwell electrodynamics. }.
 
In order to address the above mentioned issues, in the present article, based on a recent \textit{analytic technique}\footnote{This analytic scheme is known as \textit{matching method} which has been widely applied in various other occasions \cite{ref15}-\cite{ref20} including the study of Meissner effect \cite{ref39}.} \cite{ref15}-\cite{ref20} and considering the probe limit several crucial properties of holographic $ s $- wave superconductors have been investigated in the background of a $ D $ dimensional Schwarzschild AdS space time. Inspired by low energy heterotic string theory, the present work replaces the conventional Maxwell action by the Power Maxwell action\footnote{Almost identical higher curvature term appears in the the low energy limit of heterotic string theory \cite{ref43},\cite{ref46}. }. Using the standard entries in the gauge/gravity dictionary, the behavior of $ s $- wave holographic superconductors have been investigated in the presence of an external static magnetic field. It is observed that the superconducting phase disappears above a critical value of the applied magnetic field ($ B_c $). Furthermore it is observed that the value of this critical field strength ($ B_c $) increases as the mass ($ m $) of the scalar field increases, which indicates the onset a harder condensation for the higher values of scalar mass.    
 
Before going further, let us briefly mention about the organization of the paper. In section 2, the basic set up for  holographic super conductors in the background of a $ D $ dimensional Schwarzschild AdS space time has been given. In section 3, the analytical computation of the critical temperature and the condensation values have been performed employing the analytical technique developed in \cite{ref15}. Computation of the Meissner effect, based on this analytic scheme have been performed in section 4. Finally, the paper is concluded in section 5.

\section{Basic set up}
In order to carry out the analysis we first provide a basic setup for the gravity duals of higher dimensional holographic superconductors. This dual description consists of a charged hairy black hole on the background of a $ D $ dimensional Schwarzschild AdS space time. The metric of a $ D $ dimensional planar Schwarzschild AdS space time may be written as,
\begin{eqnarray}
ds^2=-f(r)dt^2+\frac{1}{f(r)}dr^2+r^2dx_{i}dx^{i}
\label{m1}
\end{eqnarray}
where,
\begin{eqnarray}
f(r)=r^{2}\left( 1-\frac{r_{+}^{D-1}}{r^{D-1}}\right)
\label{metric}
\end{eqnarray}
in units in which the AdS radius is unity, i.e. $l=1$.
The Hawking temperature for the black hole could be easily found to be,
\begin{eqnarray}
T=\left[\frac{f'(r)}{4\pi} \right]_{r=r_+}= \frac{(D-1)r_+}{4\pi}~.
\label{temp}
\end{eqnarray}

In order to carry out the analysis, we need to know the appropriate action for the hairy configuration that triggers a superconductivity at the boundary field theory. For the present case, the action for the bulk theory which includes a complex scalar field ($ \psi $) that is minimally coupled to the Power Maxwell field as\footnote{similar higher curvature ($ F^{q} $) terms also appears in the low energy limit of heterotic string theory \cite{ref43},\cite{ref46}.},
\begin{equation}
S= \int d^{D}x\sqrt{-g}\left[R-2\Lambda -\beta (F_{\mu\nu}F^{\mu\nu})^{q}-|\nabla_{\mu}\psi-iA_{\mu}\psi|^2-m^2|\psi|^2 \right], 
\end{equation}  
where $ \beta $ is the coupling constant and $ q $ is the power parameter of the Power Maxwell field. For $ \beta =1/4 $ and $ q = 1 $ the Power Maxwell action reduces to the usual Maxwell case. Here $ \Lambda\left( =-\frac{(D-1)(D-2)}{2}\right)  $ is the cosmological constant. 

In the probe limit, varying the action it is straightforward to calculate the Maxwell and scalar field equations. These may be written as,
\begin{eqnarray}
\frac{4\beta q}{\sqrt{-g}}\partial_{\mu}\left(\sqrt{-g}(F_{\lambda\sigma}F^{\lambda\sigma})^{q-1} F^{\mu\nu}\right)-i\left( \psi^{*}\partial^{\nu}\psi - \psi(\partial^{\nu}\psi)^{*}\right)-2A^{\nu}|\psi|^{2} = 0 \label{eq1}
\end{eqnarray} 
and,
\begin{eqnarray}
\partial_{\mu}\left(\sqrt{-g}\partial^{\mu}\psi \right) -i\sqrt{-g}A^{\mu}\partial_{\mu}\psi -i\partial_{\mu}\left(\sqrt{-g}A^{\mu}\psi \right)-\sqrt{-g}A^{2}\psi -\sqrt{-g}m^{2}\psi =0\label{eq2} 
\end{eqnarray}
respectively.

In order to solve the the above set of equations(\ref{eq1},\ref{eq2}) let us consider the following ansatz \cite{ref7},
\begin{eqnarray}
A=\phi(r) dt,\;\;\;\;\psi=\psi(r).
\label{vector}
\end{eqnarray}
Based on the above ansatz (\ref{vector}), the above set of equations (\ref{eq1},\ref{eq2}) turns out to be a set of radial equations which may be written as,
\begin{eqnarray}
\partial_{r}^{2}\phi+\frac{1}{r}\left( \frac{D-2}{2q-1}\right) \partial_r\phi
-\frac{2\phi\psi^2 (\partial_r\phi)^{2(1-q)}}{(-1)^{q-1}2^{q+1}\beta q (2q-1)f(r)}=0\label{eq3}
\end{eqnarray}
and,
\begin{eqnarray}
\partial_r^{2}\psi+\left(\frac{f'}{f}+\frac{D-2}{r}\right)\partial_r\psi
+\frac{\phi^{2}\psi}{f^2}-\frac{m^2\psi}{f}=0\label{eq4}
\end{eqnarray}
respectively.\\

For the future convenience let us change the variable from $ r $ to $ z(=\frac{r_+}{r}) $ so that the above set of radial equations (\ref{eq3},\ref{eq4}) eventually turn out to be a set of equations in $ z (0<z\leq1)$ which may be expressed as,
\begin{eqnarray}
\partial_{z}^{2}\phi +\frac{1}{z}\left( 2-\frac{D-2}{2q-1}\right) \partial_z\phi
+\frac{2\phi(z)\psi^2(z)r_+^{2q} (\partial_z\phi)^{2(1-q)}}{z^{4q}2^{q+1}(-1)^{3q}\beta q (2q-1)f(z)}=0\label{eq5}
\end{eqnarray}
and,
\begin{eqnarray}
\partial_z^{2}\psi+\left(\frac{f'}{f}-\frac{D-4}{z}\right)\partial_z\psi
+\frac{\phi^{2}\psi r_+^{2}}{z^{4}f^2}-\frac{m^2\psi r_+^{2}}{z^{4}f}=0\label{eq6}
\end{eqnarray}
respectively.

Before we proceed further let us first mention the boundary conditions which may be put as follows: 

 At the horizon $ z=1 $ one must have,
\begin{eqnarray}
\phi (1)=0,~~~\psi^{'}(1)=-\frac{m^{2}}{D-1}\psi(1) .\label{eq7}
\end{eqnarray}
 On the other hand, in the asymptotic AdS region ($ z\rightarrow 0 $) the solutions for (\ref{eq5},\ref{eq6}) could be expressed as,
\begin{eqnarray}
\phi(z) = \mu - \frac{\rho^{\frac{1}{2q-1}}}{r_+^{\frac{D-2}{2q-1}-1}}z^{\frac{D-2}{2q-1}-1},~~~~\psi(z)= \frac{\psi_-}{r_+^{\lambda_-}}z^{\lambda_-} + \frac{\psi_+}{r_+^{\lambda_+}}z^{\lambda_+}\label{eq8}
\end{eqnarray}
with,
\begin{equation}
\lambda_{\pm} = \frac{1}{2}[(D-1)\pm\sqrt{(D-1)^{2}+4m^{2}}].
\end{equation}  
Here $ \mu $ and $ \rho $ are the chemical potential and the charge density of the dual field theory whereas, on the other hand, $ \psi_- (= <\mathcal{O}_->) $ and $ \psi_+ (= <\mathcal{O}_+>) $ correspond to the vacuum expectation values of the dual operator $ \mathcal{O} $. In the following analysis we set $ \psi_{-}=0 $ which is a convenient boundary condition.

\section{Condensation without magnetic field}
With the above set up in hand, as a next step we derive the critical temperature ($ T_c $) as well as the critical exponent associated with the condensation values in the presence of a Power Maxwell field in higher dimensions. In order to do that, as a first step, we Taylor expand both $ \phi(z) $ and $ \psi(z) $ near the horizon as,
\begin{eqnarray}
\phi (z)=\phi (1) - \phi^{'}(1)(1-z) + \frac{1}{2}\phi^{''}(1)(1-z)^{2} + O((1-z)^{3})\label{eq9}
\end{eqnarray}
and,
\begin{eqnarray}
\psi(z)=\psi(1)-\psi^{'}(1)(1-z)+\frac{1}{2} \psi^{''}(1)(1-z)^{2} + O ((1-z)^{3})\label{eq10}
\end{eqnarray}
respectively, where the prime corresponds to the derivative w.r.t $ z $. Also, we choose $ \phi'(1)<0 $ and $ \psi(1)>0 $ with out loss of generality.

As a next step, we note that for $ z=1 $  from (\ref{eq5}) we find,
\begin{eqnarray}
\phi^{''}(1)&=& -\left[\frac{1}{z}\left( 2-\frac{D-2}{2q-1}\right) \partial_z\phi \right]_{z=1}-\left[\frac{2\phi(z)\psi^2(z)r_+^{2q} (\partial_z\phi)^{2(1-q)}}{z^{4q}2^{q+1}(-1)^{3q}\beta q (2q-1)f(z)} \right]_{z=1}\nonumber\\
&=& - \left(2 - \frac{D-2}{2q-1} \right)\phi'(1) + \frac{2r_+^{2(q-1)}\psi^{2}(1)(\phi'(1))^{3-2q}}{(-1)^{3q} 2^{q+1}\beta q (2q-1)(D-1)} \label{eq11}
\end{eqnarray}
where we have used the Leibnitz rule in order to obtain the above result. Finally, substituting (\ref{eq11}) into (\ref{eq9}) we obtain,
\begin{equation}
\phi(z)= -\phi'(1)(1-z) + \frac{1}{2}\left[ \left( \frac{D-2}{2q-1} -2 \right)\phi'(1) + \frac{2r_+^{2(q-1)}\psi^{2}(1)(\phi'(1))^{3-2q}}{(-1)^{3q} 2^{q+1}\beta q (2q-1)(D-1)} \right](1-z)^{2} \label{eq12}.
\end{equation}
On the other hand, following similar steps as mentioned above and using (\ref{eq7}) from (\ref{eq6}) near $ z=1 $ we find,
\begin{equation}
\psi''(1)= -\frac{m^{2}}{(D-1)}\left( 1 - \frac{m^{2}}{2(D-1)}\right)\psi(1) - \frac{\phi'^{2}(1)\psi(1)}{2r_+^{2}(D-1)^{2}} \label{eq13}.
\end{equation} 
Substituting (\ref{eq13}) into (\ref{eq10}) we finally obtain,
\begin{eqnarray}
\psi (z) = \left( 1 + \frac{m^{2}}{D-1}\right)\psi(1)- \frac{m^{2}}{D-1}\psi(1)z\nonumber\\
-\frac{1}{2}\left[\frac{m^{2}}{(D-1)}\left( 1 - \frac{m^{2}}{2(D-1)}\right)+ \frac{\phi'^{2}(1)}{2r_+^{2}(D-1)^{2}} \right]\psi(1)(1-z)^{2} \label{eq14}. 
\end{eqnarray}

With the above expressions in hand, and keeping terms only upto quadratic order in the Taylor expansion (\ref{eq9},\ref{eq10}) it is now quite straightforward to obtain an analytic expression for the critical temperature ($ T_c $) and the corresponding condensation values by matching the solutions (\ref{eq8}), (\ref{eq12}) and (\ref{eq14}) at some intermediate point $ z=z_m $ \cite{ref15}. This finally yields the following set of equations,
\begin{eqnarray}
\mu -\frac{\rho^{\frac{1}{2q-1}}z_m^{\frac{D-2}{2q-1}-1}}{(r_+)^{\frac{D-2}{2q-1}-1}} \approx \frac{\vartheta}{2}(1 - z_m)\left[ 2+\left( 2 -\frac{D-2}{(2q-1)}\right) (1-z_m) \right]\nonumber\\  + \frac{r_+^{2(q-1)}\alpha^{2}(-\vartheta)^{3-2q}(1- z_m)^{2}}{(-1)^{3q}2^{q+1}\beta q(2q-1)(D-1)} \label{eq15}
\end{eqnarray}
\begin{eqnarray}
-\frac{\rho^{\frac{1}{2q-1}}z_m^{\frac{D-2}{2q-1}-2}}{(r_+)^{\frac{D-2}{2q-1}-1}}\left(\frac{D-2}{2q-1}-1\right)  \approx -\vartheta\left[ 1+ \left(2 - \frac{D-2}{2q-1}\right)(1-z_m)\right] \nonumber\\ - \frac{2 r_+^{2(q-1)}\alpha^{2}(-\vartheta)^{3-2q}(1-z_m)}{(-1)^{3q} 2^{q+1}\beta q (2q-1)(D-1)} \label{eq16}
\end{eqnarray}
\begin{eqnarray}
\frac{<\mathcal{O}_+> z_m^{\lambda_+}}{r_+^{\lambda_+}} \approx \alpha\left( 1+\frac{m^{2}(1-z_m)}{(D-1)}\right)\nonumber\\
-\frac{\alpha}{2}\left(\frac{m^{2}}{D-1}\left( 1-\frac{m^{2}}{2(D-1)}\right) +\frac{\tilde{\vartheta}^{2}}{2(D-1)^{2}} \right)(1-z_m)^{2} \label{eq17} 
\end{eqnarray}
and,
\begin{equation}
\frac{\lambda_+<\mathcal{O}_+>z_m^{\lambda_+-1}}{r_+^{\lambda_+}}  \approx -\frac{m^{2}\alpha}{D-1} + \alpha \left(\frac{m^{2}}{D-1}-\frac{m^{4}}{2(D-1)^{2}}+\frac{\tilde{\vartheta}^{2}}{2(D-1)^{2}} \right)(1-z_m) \label{eq18}
\end{equation}
where we have set $ \vartheta=-\phi'(1) $, $ \alpha = \psi(1) $ and $ \tilde{\vartheta}=\frac{\vartheta}{r_+} $.

At this stage our aim is to calculate the unknown entities $ \tilde{\vartheta} $ and $ <\mathcal{O}_+> $ in terms of temperature ($ T $) and other known parameters of the theory. Considering (\ref{eq16}) and using (\ref{temp}) it is quite straightforward to obtain,
\begin{eqnarray}
\alpha^{2}=\frac{(-1)^{5q-3}2^{q}\beta q (2q-1)(D-1)}{\tilde{\vartheta}^{2(1-q)}(1-z_m)}\left[ 1+\left( 2-\frac{D-2}{2q-1}\right)(1-z_m)\right]\nonumber\\ \times \left(\frac{T_c}{T} \right)^{\frac{D-2}{2q-1}}\left[1- \left(\frac{T}{T_c} \right)^{\frac{D-2}{2q-1}}\right] \label{eq20},  
\end{eqnarray}
where,
\begin{equation}
T_c = \gamma \rho^{\frac{1}{D-2}}\label{eq21}
\end{equation}
with,
\begin{equation}
\gamma = \frac{z_m^{\frac{(\frac{D-2}{2q-1}-2)(2q-1)}{D-2}}}{\tilde{\vartheta}^{\frac{2q-1}{D-2}}}\left( \frac{D-1}{4\pi}\right)\frac{\left(\frac{D-2}{2q-1}-1 \right)^{\frac{2q-1}{D-2}} }{\left[ 1+\left(2-\frac{D-2}{2q-1} \right)(1-z_m) \right]^{\frac{2q-1}{D-2}}}.
\end{equation}

From (\ref{eq17}) and (\ref{eq18}) one can further obtain,
\begin{equation}
\tilde{\vartheta} = \sqrt{m^{4}+2m^{2}(D-1)\left[ \frac{2(z_m+\lambda_+(1-z_m))}{(1-z_m)(2z_m+\lambda_+(1-z_m))}-1\right]  +\frac{4\lambda_{+}(D-1)^{2}}{(1-z_m)(2z_m+\lambda_{+}(1-z_m))}}
\end{equation}
and,
\begin{equation}
<\mathcal{O}_{+}> = \frac{r_+^{\lambda_{+}}}{z_m^{\lambda_{+}}}\frac{\left(1+\frac{m^{2}(1-z_m)}{2(D-1)} \right) }{\left( 1+\frac{\lambda_{+}(1-z_m)}{2z_m}\right) }\sqrt{\mathcal{A}}\left(\frac{T_c}{T} \right)^{\frac{D-2}{2(2q-1)}}\sqrt{1-\left( \frac{T}{T_c}\right)^{\frac{D-2}{2q-1}} }  \label{eq23}
\end{equation}
with,
\begin{equation}
\mathcal{A} = \frac{(-1)^{5q-3}2^{q}\beta q (2q-1)(D-1)}{\tilde{\vartheta}^{2(1-q)}(1-z_m)}\left[ 1+\left( 2-\frac{D-2}{2q-1}\right)(1-z_m)\right].
\end{equation}
\begin{table}[htb]
\caption{Comparison between the analytical and numerical values}   
\centering                          
\begin{tabular}{c c c c c c c c}            
\hline\hline                        
$ z_m $& $D$ & $ q $& $ m^{2} $ & $\gamma_{matching}$  & $\gamma_{numerical}$ &  \\ [0.05ex]
\hline
0.5&5 & 1 & 0 & 0.1702& 0.1700 \\
0.5&5 & 1 & -1& 0.1741& 0.1765  \\
0.5&5 & 1 & -2& 0.1783 &0.1847  \\ [0.5ex] 
\hline                              
\end{tabular}\label{E1}  
\end{table}
\begin{table}[htb]
\caption{Comparison between the analytical and numerical values}   
\centering                          
\begin{tabular}{c c c c c c c c}            
\hline\hline                        
$ z_m $& $D$ & $ q $& $ m^{2} $ & $\gamma_{matching}$  & $\gamma_{numerical}$ &  \\ [0.05ex]
\hline
0.5&5 & $\frac{5}{4}$ & 0 &0.0880& 0.1008 \\
0.5&5 & $\frac{5}{4}$ & -1& 0.0910& 0.1065  \\
0.5&5 & $\frac{5}{4}$ & -2& 0.0944&0.1145  \\ [0.5ex] 
\hline                              
\end{tabular}\label{E1}  
\end{table}

Before we proceed further, let us note few points at this stage. First of all, (from table 1 and table 2) we note that both the analytic and numerical results \cite{ref28} are in quite good agreement with each other. This indeed ensures the validity of the matching scheme. Also we note that the proportionality coefficient ($ \gamma $) decreases for higher values of the scalar mass ($ m $) and Power parameter ($ q $) which indicates the onset of a harder condensation (see fig 1). Moreover, from (\ref{eq23}) we note that the critical exponent associated with the condensation near the critical point is $\frac{1}{2}$ which is the universal feature of mean field theory.

\section{Meissner like effect}

With the above expressions in hand, we now aim to investigate the nature of the condensate solutions in presence of an external magnetic field. Furthermore, we also investigate how the condensate at low temperatures is affected due to the presence of scalar mass ($ m $). In order to do that, we need some extra piece in our theory, namely adding an external magnetic field in the bulk theory. According to the gauge/gravity dictionary, the asymptotic value of this magnetic field corresponds to a magnetic field added to the boundary field theory, i.e, $ B({\bf{x}})=F_{xy}({\bf{x}},z\rightarrow 0) $.  It is obvious that the condensate will be very small everywhere near the critical field strength  $ B\sim B_{c} $. Therefore we may regard $ \psi (x,z)$ as a perturbation in the corresponding dual theory. Based on the above physical arguments, we adopt the following ansatz \cite{ref37},
\begin{equation}
A_t = \phi(z),~~~A_y = B x,~~~and~~~\psi = \psi (x,z)\label{eq24}.
\end{equation}

With the above choice, the scalar field equation (\ref{eq2}) for $ \psi $ turns out to be,
\begin{eqnarray}
\partial_z^{2}\psi(x,z)+ \left( \frac{f^{'}(z)}{f(z)} - \frac{D-4}{z}\right) \partial_z\psi(x,z)+ \frac{r_+^{2}\phi^{2}(z)\psi(x,z)}{z^{4}f^{2}(z)} - \frac{m^{2}r_+^{2}\psi(x,z)}{z^{4}f(z)}\nonumber\\
+\frac{1}{z^{2}f(z)}(\partial_{x}^{2}\psi - B^{2}x^{2}\psi)=0\label{eq25}.
\end{eqnarray}
 
In order to solve (\ref{eq25}), we consider the following separable form
\begin{equation}
\psi (x,z) = X(x) R(z)\label{eq26}.
\end{equation}
 
 Substituting (\ref{eq26}) into (\ref{eq25}) one can easily obtain the following,
 \begin{eqnarray}
\frac{ z^{2}f(z)}{R(z)}\left[\partial_z^{2}R(z)+ \left( \frac{f'}{f}-\frac{D-4}{z}\right)\partial_z R(z)\right] +\frac{\phi^{2}(z)r_+^{2}}{z^{2}f(z)}-\frac{m^{2}r_{+}^{2}}{z^{2}} \nonumber\\
-\frac{1}{X(x)}\left[ -\partial_x^{2} X(x) + B^{2} x^{2} X(x)\right]=0.  
 \end{eqnarray}
 
 Note that $ X(x)$ satisfies Schrodinger equation for a simple harmonic oscillator localized in one dimension with frequency determined by $ B $ \cite{ref6,ref37,ref39},
 \begin{equation}
 -X^{''}(x) + B^{2}x^{2}X(x) = \lambda_{n} B X(x)
 \end{equation}
 where $ \lambda_{n}= 2n+1 $ denotes the separation constant. We take the lowest mode ($ n=0 $) solution for the rest of our analysis as it is the most stable one \cite{ref6},\cite{ref39}.  
 
 With this particular choice, the corresponding equation for $ R(z) $ takes the following form,
 \begin{equation}
 R''(z)+\left( \frac{f^{'}(z)}{f(z)}- \frac{D-4}{z}\right) R'(z)+ \frac{r_+^{2}\phi^{2}(z)R(z)}{z^{4}f^{2}(z)}-\frac{m^{2}r_+^{2}R(z)}{z^{4}f(z)}=\frac{B R(z)}{z^{2}f(z)}\label{eq27}.
 \end{equation}
 
 Following our previous approach, we expand $ R(z) $ near the horizon ($ z=1 $) in a Taylor series as \cite{ref39},
  \begin{equation}
 R(z) = R (1) - R^{'}(1)(1-z) + \frac{1}{2} R^{''}(1)(1-z)^{2} + O ((1-z)^{3})\label{eq30}
 \end{equation}
 
 On the other hand, the asymptotic ($ z\rightarrow 0 $) behavior of $ R(z)$ could be written as, 
\begin{equation}
 R(z) =  \frac{<\mathcal{O}_+>}{r_+^{\lambda_+}} z^{\lambda_+} \label{eq29}
 \end{equation} 
where according to our previous choice $ <\mathcal{O}_{-}> = 0 $.

Following the arguments of matching technique, we match (\ref{eq30}) and (\ref{eq29}) at some intermediate point $ z=z_m $ which may be put as,
\begin{eqnarray}
\left[ \frac{<\mathcal{O}_+>}{r_+^{\lambda_+}} z^{\lambda_+} \right]_{z=z_m}=\left[  R (1) - R^{'}(1)(1-z) + \frac{1}{2} R^{''}(1)(1-z)^{2} + O((1-z)^{3})\right]_{z=z_m}\label{r}  
\end{eqnarray}

In order to proceed further, we need to calculate $ R'(1) $ and $ R''(1) $. From (\ref{eq27}) we find,
 \begin{equation}
 R'(1) = \left(\frac{-m^{2}}{D-1} - \frac{B}{r_+^{2}(D-1)}\right) R(1)\label{eq28}. 
\end{equation}
 
On the other hand, from (\ref{eq27}) and using (\ref{eq28}) we find,
\begin{equation}
R''(1) = \left[\left(1-\frac{m^{2}}{2(D-1)}-\frac{B}{2r_+^{2}(D-1)} \right)\left(\frac{-m^{2}}{D-1} - \frac{B}{r_+^{2}(D-1)} \right)-\frac{\phi'^{2}(1)}{2r_+^{2}(D-1)^{2}}  \right]R(1)\label{eq31}. 
\end{equation} 
 Substituting (\ref{eq28},\ref{eq31}) into (\ref{r}) and setting $ z = z_m $ we find,
\begin{eqnarray}
\frac{z_m^{\lambda_+}<\mathcal{O}_+>}{r_+^{\lambda_+}} \approx \left( 1+\frac{m^{2}(1-z_m)}{(D-1)}\right)R(1) + \frac{BR(1)(1-z_m)}{r_+^{2}(D-1)}\nonumber\\
+\frac{1}{2}\left[\left(1-\frac{m^{2}}{2(D-1)}-\frac{B}{2r_+^{2}(D-1)} \right)\left(\frac{-m^{2}}{D-1} - \frac{B}{r_+^{2}(D-1)} \right)-\frac{\phi'^{2}(1)}{2r_+^{2}(D-1)^{2}}  \right]R(1)(1-z_m)^{2}\label{eq32}. 
\end{eqnarray}
As a next step, we differentiate (\ref{r}) w.r.t $ z $ and again set $ z=z_m $, which finally yields,
\begin{eqnarray}
\lambda_{+}\frac{z_m^{\lambda_+ -1}<\mathcal{O}_+>}{r_+^{\lambda_+}} \approx -\frac{m^{2}}{(D-1)}R(1) - \frac{BR(1)}{r_+^{2}(D-1)}\nonumber\\
-\left[\left(1-\frac{m^{2}}{2(D-1)}-\frac{B}{2r_+^{2}(D-1)} \right)\left(\frac{-m^{2}}{D-1} - \frac{B}{r_+^{2}(D-1)} \right)-\frac{\phi'^{2}(1)}{2r_+^{2}(D-1)^{2}}  \right]R(1)(1-z_m)\label{eq33}.
\end{eqnarray}

Using the above two equations (\ref{eq32},\ref{eq33}), it is now quite straightforward to calculate the expression for the magnetic field ($ B $). After some algebraic steps, we arrive at the following quadratic equation in $ B $,
\begin{equation}
B^{2} + pr_+^{2}B+nr_+^{4}-\phi'^{2}(1)r_+^{2}=0
\end{equation}
where,
\begin{equation}
p = 2m^{2}-2(D-1)+\frac{4(D-1)(\lambda_+(1-z_m) +z_m)}{(1-z_m)(\lambda_{+}(1-z_m)+2z_m)}
\end{equation}
and,
\begin{equation}
n=m^{4}-2m^{2}(D-1)+\frac{2(D-1)^{2}\left( \frac{\lambda_{+}}{z_m}\left( 1+\frac{m^{2}(1-z_m)}{D-1}\right)+\frac{m^{2}}{D-1}\right)  }{(1-z_m)\left(\frac{\lambda_+(1-z_m)}{2z_m} +1\right) }. 
\end{equation}

Let us now consider the case $ B\sim B_{c} $, which implies that the condensation is very small ($ \psi\sim 0 $) everywhere and therefore we may ignore all the higher order terms in $ \psi $. With this above argument, the corresponding equation in $ \phi(z) $ turns out to be,
\begin{eqnarray}
\partial_{z}^{2}\phi +\frac{1}{z}\left( 2-\frac{D-2}{2q-1}\right) \partial_z\phi \approx 0
\end{eqnarray}
which has a unique solution,
\begin{eqnarray}
\phi(z)&=&\left(\frac{\rho}{r_+^{D-2}} \right)^{\frac{1}{2q-1}} r_+(1-z^{\frac{D-2}{2q-1}-1})\nonumber\\
\Rightarrow \phi'^{2}(1)r_+^{2}&=&\frac{\rho^{2/(2q-1)}}{r_+^{\frac{2(D-2)}{2q-1}}}r_+^{4}\left( \frac{D-2}{2q-1}-1\right)^{2}\label{eq34}. 
\end{eqnarray}
\begin{figure}[h]
\centering
\includegraphics[angle=0,width=16cm,keepaspectratio]{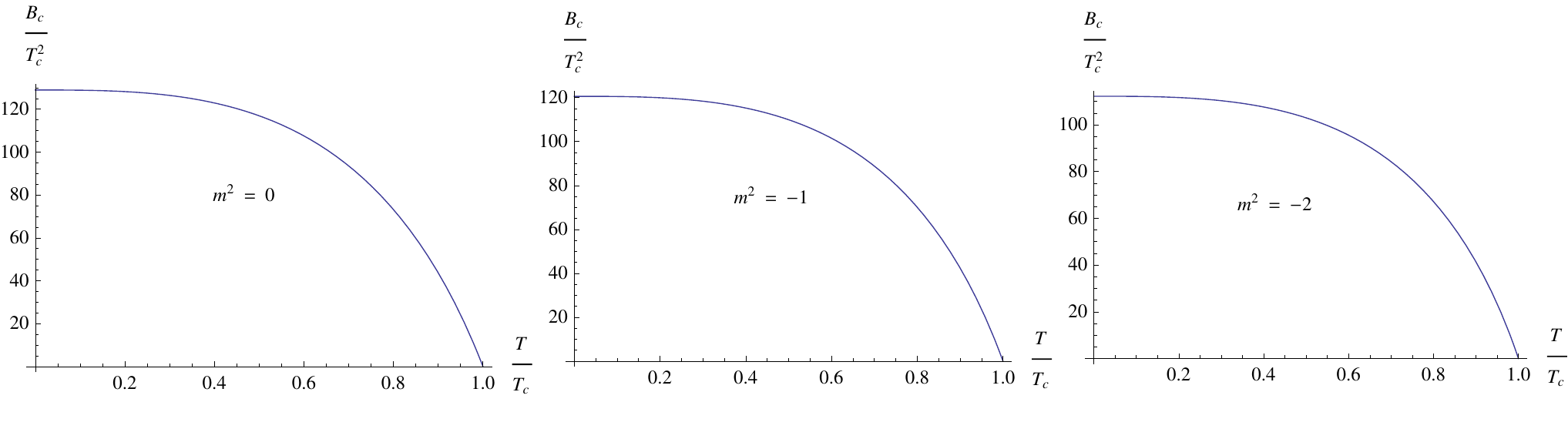}
\caption[]{\it Critical magnetic field ($ B_{c}$) plot for holographic superconductors with $ z_m =0.5 $, $ D=5 $ and $ q=1 $.}
\label{figure 2a}
\end{figure}
\begin{figure}[h]
\centering
\includegraphics[angle=0,width=16cm,keepaspectratio]{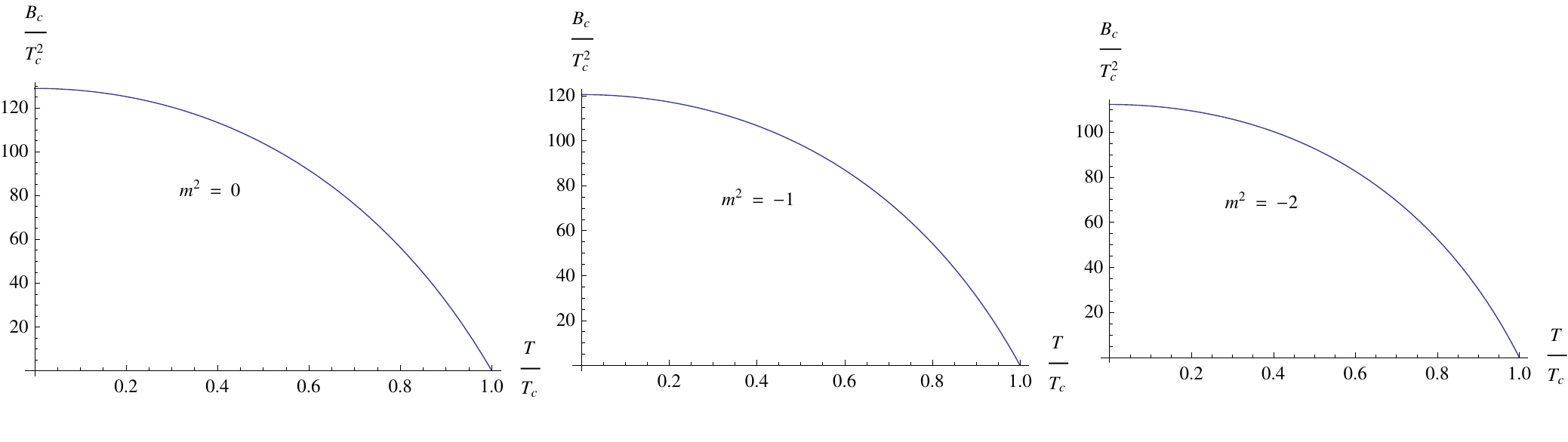}
\caption[]{\it Critical magnetic field ($ B_{c}$) plot for holographic superconductors with $ z_m =0.5 $, $ D=5 $ and $ q=5/4 $.}
\label{figure 2a}
\end{figure}

Using (\ref{temp}), (\ref{eq21}) and (\ref{eq34}) after some algebraic steps we finally obtain,
\begin{eqnarray}
B_{c}=\frac{(D-1)^{\frac{D-2}{2q-1}-2}}{2(4\pi)^{\frac{D-2}{2q-1}-2}\gamma^{\frac{D-2}{2q-1}}} T_c^{2}\left[ \Omega (D,q,m)-p\left( \frac{4\pi\gamma}{D-1}\right)^{\frac{(D-2)}{(2q-1)}}\left(\frac{T}{T_c} \right)^{\frac{(D-2)}{(2q-1)}}\right]\label{bcfinal} 
\end{eqnarray}
where,
\begin{equation}
\Omega (D,q,m)=\sqrt{4\left(\frac{D-2}{2q-1}-1 \right)^{2}-(4n-p^{2})\left( \frac{4\pi\gamma}{D-1}\right)^{\frac{2(D-2)}{(2q-1)}}\left(\frac{T}{T_c} \right)^{\frac{2(D-2)}{(2q-1)}}}.
\end{equation}
 
 From the above figures (fig. 1 and fig. 2) it is indeed evident that there exists a critical magnetic field ($ B_c $) above which the superconductivity ceases to exist. The behavior of this critical magnetic field with temperature ($ T $) is quite similar to that is observed in ordinary type II superconductors \cite{ref33}-\cite{ref34}. Interestingly, similar behavior is also predicted by the Ginzburg-Landau theory where an expression like (\ref{bcfinal}) could be derived for ordinary superconductors \cite{ref33}-\cite{ref34}.  This seems to be a unique qualitative feature of holographic superconductors which is valid even in higher dimensions. Furthermore, we observe that the situation does not alter even when we incorporate non linear corrections ($ q>1 $) to the usual Maxwell action. Finally, and most importantly we note that the condensation at low temperatures is indeed affected due to the presence of the scalar mass even in the presence of external magnetic field. The upper critical field strength strength ($ B_{c} $) has been found to be higher for the larger mass ($ m $) of the scalar particles which suggests the onset of a harder condensation.


\section{Conclusions}
 
 In this paper, considering the probe limit various properties of holographic superconductors have been investigated using the matching technique \cite{ref15}-\cite{ref16}. The present paper considers holographic superconductors immersed in an external magnetic field in the presence Power Maxwell corrections to the usual Maxwell action. Such higher curvature corrections are found to be emerging naturally in the low energy limit of heterotic string theory \cite{ref43}-\cite{ref46}.  The critical temperature ($ T_c $) has been found to be suppressed for the higher values of the power parameter ($ q $). Moreover, the critical exponent associated with the condensation value near the critical point has been found to be equal to $ 1/2 $ which is in good agreement with the universal mean field value and indeed suggests the onset of a second order phase transition.  

Finally, it is observed that holographic superconductors when immersed in an external static magnetic field, can support a larger magnetic filed at low temperatures which could be observed even in higher dimensions. It is observed that the superconducting phase exists only below a critical field strength ($ B_c $). Interestingly similar behavior could be found in ordinary type II superconductors, where the strength of the upper critical magnetic field ($ B_{c2} $) has been found to be increasing as the temperatures is lowered through $ T_c $ \cite{ref33}-\cite{ref34}. Most importantly we note that the condensation also gets affected due to the presence of the scalar mass ($ m $) in the theory. It is observed that the critical value of the magnetic field ($ B_c $) increases as $ m^{2} $ shifts towards a larger value. The magnetic field as it grows will try to squeeze the condensate away completely. This indeed suggests the fact that the scalar hair formation at low temperatures ($ T<T_c $) gets more difficult for the larger mass of the scalar particles.  


\vskip 5mm
{\bf{Acknowledgement :}}\\
     Author would like to thank the Council of Scientific and Industrial Research (C. S. I. R), Government of India, for financial help.


\end{document}